# The State of Infodemic on Twitter

Drishti Jain[*], Tavpritesh Sethi[*]

[*] Indraprastha Institute of Information Technology

*Abstract*- Following the wave of misinterpreted, manipulated and malicious information growing on the Internet, the misinformation surrounding COVID-19 has become a paramount issue. In the context of the current COVID-19 pandemic, social media posts and platforms are at risk of rumors and misinformation in the face of the serious uncertainty surrounding the virus itself. At the same time, the uncertainty and new nature of COVID-19 means that other unconfirmed information that may appear "rumored" may be an important indicator of the behavior and impact of this new virus. Twitter, in particular, has taken a center stage in this storm where Covid-19 has been a much talked about subject. We have presented an exploratory analysis of the tweets and the users who are involved in spreading misinformation and then delved into machine learning models and natural language processing techniques to identify if a tweet contains misinformation.

*Index Terms*- Covid-19, Infodemic, Misinformation, Twitter

## I. INTRODUCTION

Even before the official announcement of the COVID-19 coronavirus infection pandemic on March 11, 2020, conspiracy theories and misinformation about the origin, scale, prevention, treatment and other aspects of this disease began to spread on the Internet. Conspiracy theories were circulated on social networks, text messages, and a number of state media outlets in various countries. Among the most common versions of this kind are claims that the virus is a biological weapon with a patented vaccine, a tool for population regulation or the result of a spy operation[1]. Much of the misinformation associated with COVID-19 involves "various forms of reconfiguration where existing and often true information is distorted, distorted, reconstructed, or rewritten," according to a study published by the Reuters Institute for Journalism Research, it is rare for misinformation to be "totally fabricated"[2]. Medical misinformation about the methods of prevention, treatment and self-diagnosis of coronavirus disease has also spread on social networks [3].

The World Health Organization has declared an "infodemic" of incorrect information about the virus, which poses risks to global health[4]

Studies have shown that many people connect to the Internet and social media platforms such as Twitter, Facebook, Whatsapp, Instagram and Reddit every day and utilize it for getting information/news through them[5]. Twitter users, in particular, are known for sharing and consuming news: 59% of Twitter users describe it as good or extremely good for sharing preventive health information[6].Clearly, people seek COVID-19 information online that they perceive to be helpful, leading to a wide range of fake news consumption and sharing[7].

The proliferation of social media has opened up exciting avenues with the potential, at the very least, to increase clarity and democracy in the sharing of scientific data .In the same way, it dramatically increased the level of trust in personal opinions (beliefs, evaluations, etc.) and allowed them to spread faster. It is clear that - information and misinformation will largely determine how people understand and respond to public health crises, as well as how they assess which institutions are helping to solve them. As information about COVID-19 evolves every day, the public is faced with a combination of partial information, conflicting information, and sometimes complete misinformation. Therefore, it is very important that people have access to reliable news and information that can help them understand this crisis, what they can do to protect themselves and society as a whole.

In the context of this acute uncertainty, we have investigated the content of misinformation on Twitter related to the topic of COVID-19 in order to achieve deeper understanding and then build a pipeline around that data to ascertain if a tweet contains misinformation or not.

We have used a corpus of 30000 tweets out of which 15000 tweets contain misinformation while the other 15000 have factual information. Here, the term misinformation refers to information that is false. In this study, we do not claim to study the intent of the user and thus, all information which is fake is termed misinformation regardless of the purveyor's intent.

The rest of the paper is structured as follows. In the next section, we discuss the data used in our work. In Section III, we describe the method and the features.This section will also analyze the impact of different attributes on misinformation. The machine learning approach used and the results of the system are explained in section IV. Finally, in Section V, the paper is concluded.

## II. DATASET

CoAID (Covid-19 heAlthcare mIsinformation Dataset)[8] is a "diverse COVID-19 healthcare misinformation dataset, including fake news on websites and social platforms, along with users' social engagement about such news". Titles of news articles were used as search queries for the Twitter API, and tweets discussing the news in question in a certain period were thus obtained. The retrieved user engagement features include: user ID, tweets, replies, favorites, retweets, and location. Consequently, the tweets were tagged based on whether they contained factual or misleading news. A second source for data was a crowd-sourced database[9] of tweets annotated in English and Arabic with fine-grained labels related to misinformation about COVID-19. The labels answer seven different questions that are of interest to journalists, fact-checkers, social media platforms, policymakers, and society as a whole. Here, only those tweets have been used which are annotated in English and which are marked as containing verifiable and factual information or which are certainly false. Tweets which only cater to questions like - "Will the tweet's claim have an effect on or be of interest to the general public?", "Is the tweet harmful for society and why?" or "Do you think that this tweet should get the attention of a government entity?" have been discarded.

III. METHOD

*A. Pre-processing and feature collection*

To identify the features in context to a tweet[11] and its user[12], the Twitter API was used. From the tweet object, the metadata of interest includes:
- Tweet id
- Source of tweet
- Tweet content
- Number of retweets
- Number of likes
- Does the tweet contain content or media identified as sensitive?
- Date the tweet was posted

Using these base attributes, we derived more attributes which could not be accessed from the Twitter API directly. We used standard natural language processing (NLP) techniques in this step. These derived attributes help us to gain more insight in the tweet parameters.
- Length of the tweet
- Question marks in the tweet
- Exclamation marks in the tweet
- Hashtags
- Number of sentences in the tweet
- Average word length
- Polarity score
- Number of uppercase characters used in the tweet
- If the tweet contains a URL
- Number of misspelled words in the tweet, if any
- POS tagging[13]

From the user object, the basic attributes collected include:
- User id
- Name
- The screen name, handle, or alias that this user identifies themselves with.
- Is the profile geo enabled?
- Bio description
- URL provided by the user in association with their profile
- Number of followers
- Number of accounts following
- The UTC datetime that the user account was created on Twitter.
- Number of Tweets this user has liked in the account's lifetime
- Number of Tweets (including retweets) issued by the user

It must be noted that a lot of tweets and users in both the datasets have been deleted after their publication, rendering them unusable and hence, they too were filtered out. Once the relevant tweet id and its label (true/false) were obtained, a Twitter API library was used to scrape the Tweet[10] and the user data. In the end, there were 15,000 tweets containing misinformation and around 1.5 lakh tweets containing real information. Since, this imbalance of data can cause aberrant behavior, we use 30000 tweets out of which 15000 contain malicious or misleading information and 15000 contain trustworthy information.

- Verified account
- Default profile image

Adding on to the user metadata features, the new features mined were:
- Length of bio
- Length of screen name
- Length of name
- Number of days since the account was created
- Number of matching characters in name and username
- Ratio of followers to following
- Ratio of tweets to days since creation
- Ratio of likes to days since creation

After cleaning and collecting all the required data, there were 15000 fake tweets and 15000 tweets with factual information. This is distributed between 29986 twitter users.

*B. Observations*

In this section, we will perform an exploratory data analysis of all the collected attributes and see how they impact the spread of misinformation.

From fig.1, it is clear that false information spreads more rapidly on the social network Twitter than real news does. It is also more likely to be retweeted or liked by the users. As observed from fig.3 tweets with false information may use excessive capital letters to emphasise or exaggerate their point and to garner attention. In the past, it has been said that people with conservative viewpoints have increasingly used frequent capitalization[14]. When it comes to measuring the readability of tweets, we could not observe much difference between tweets with false information and tweets professing the truth. Fig.4, fig.5 and fig.6 exhibit our observation using three different tests. Readability scores are formulas for assessing the readability of text, usually by counting syllables, words, and sentences. The Flesch–Kincaid ease[15] is readability tests designed to indicate how difficult a passage in English is to understand.

$$206.835 - 1.015 \left( \frac{\text{total words}}{\text{total sentences}} \right) - 84.6 \left( \frac{\text{total syllables}}{\text{total words}} \right)$$

The SMOG grade[16] is a measure of readability that estimates the years of education needed to understand a piece of writing.

$$\text{grade} = 1.0430 \sqrt{\text{number of polysyllables} \times \frac{30}{\text{number of sentences}}} + 3.1291$$

Unlike the other indices, the automated readability index (ARI)[17], relies on a factor of characters per word, instead of the usual syllables per word.

$$4.71 \left( \frac{characters}{words} \right) + 0.5 \left( \frac{words}{sentences} \right) - 21.43$$

Despite using different measures, there seems to be little that distinguishes between real and false when judging the ease with which a reader can understand a written text. We believe this may be due to the extensive use of Twitter Slang, lingo, Twitter abbreviations and acronyms. Apart from that, with only 280 characters to convey your thoughts, users often misspell words and skip syncategorematic terms (articles, connectives, prepositions, quantifiers). While fig.7 refers to the number of days since a particular account under consideration was created, Fig.8 refers to the number of exclamation marks posted by the user in a tweet. From these graphs, we can see that there is not much difference between users posting real news and false tweets based on these parameters alone. However, in Fig.9, it is clearly visible that the ratio of followers and friends produces different behaviour between the two. One hypothesis could be that bots which are more prone to posting misinformation are more likely to have few followers. In the same vein, this could also be attributed to the rapid liking in a short interval. Fig. 11 denotes the graph for polarity of the content of the tweet. Polarity has a floating value in [-1,1] where the positive values tend to a positive statement and the negative values hint towards a negative statement. From our dataset, 16151 tweets had a neutral outlook, 9977 tweets hinted towards a positive outlook and the remaining 3872 tweets had a negative polarity associated with them. Fig 18. shows wordclouds associated with all the three sentiments. That being said, we also observed that parameters like length of username, length of user bio/description, number of matching characters in the two, number of sentences in the tweet or average word length do not give away much useful information.

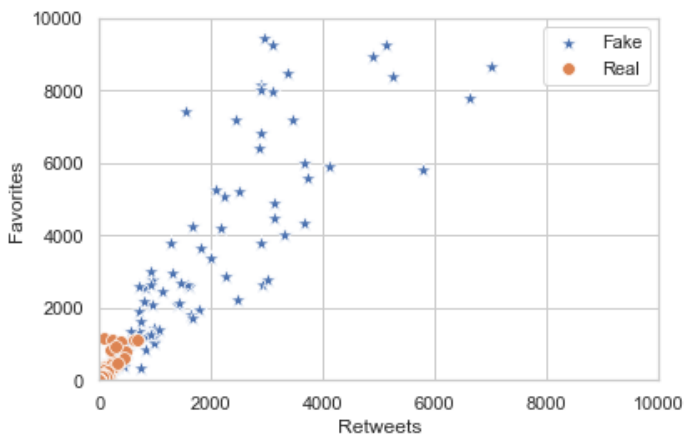

Fig. 1

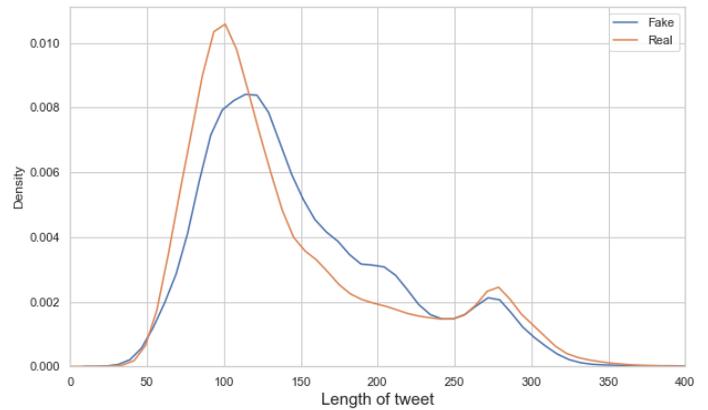

Fig 2.

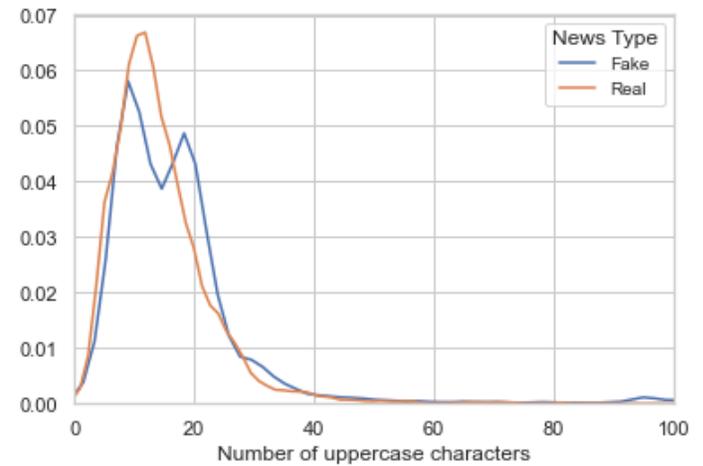

Fig. 3

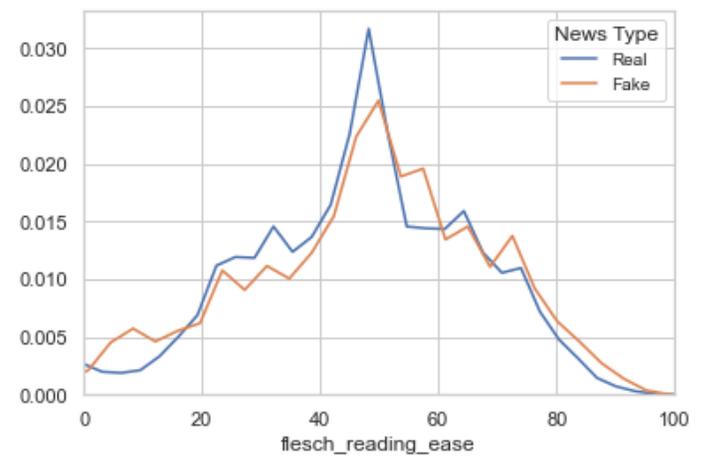

Fig.4

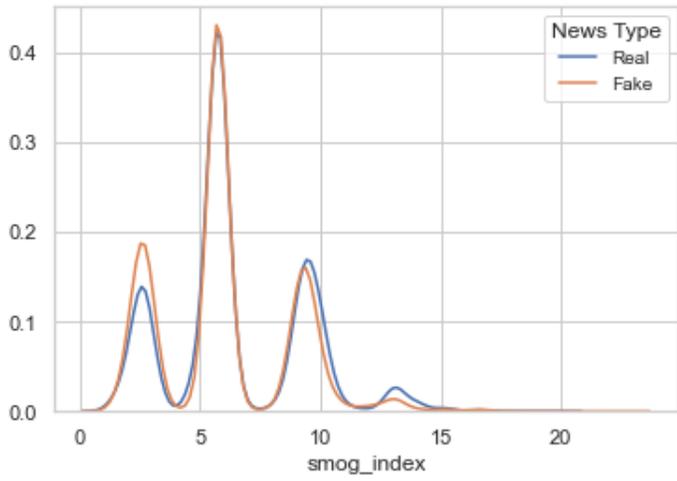

Fig.5

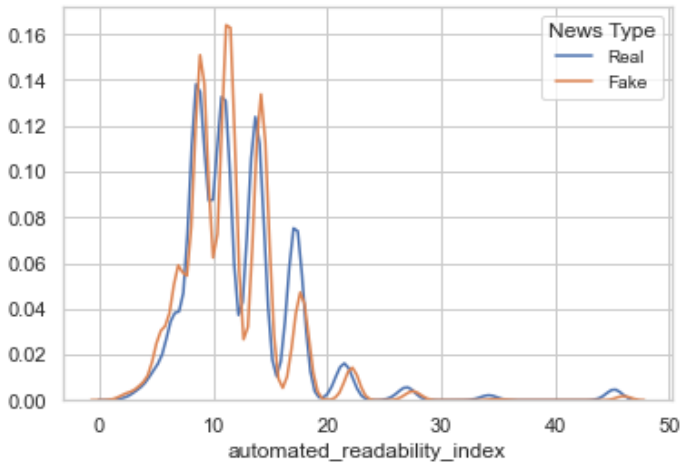

Fig.6

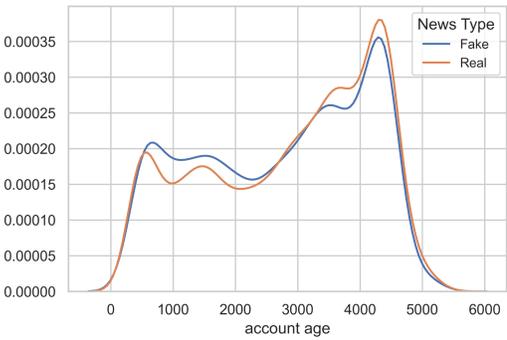

Fig. 7

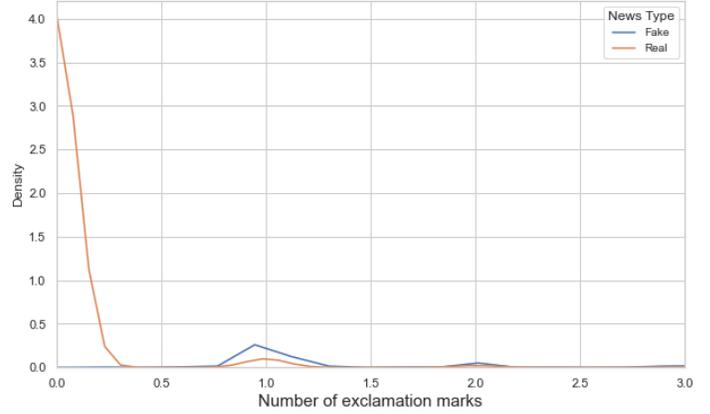

Fig 8

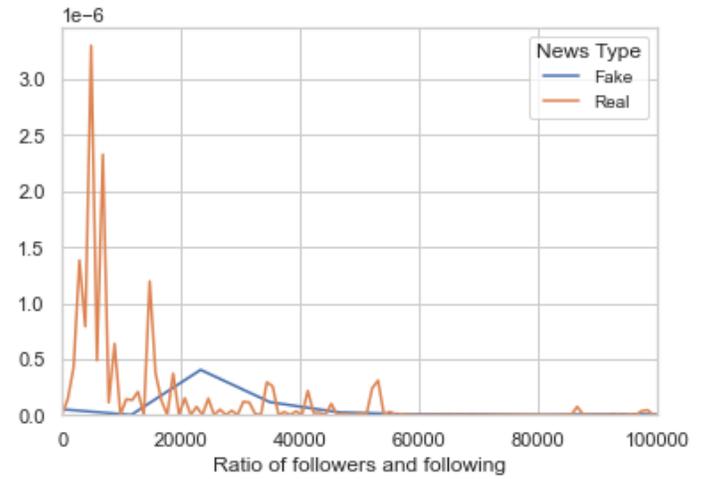

Fig.9

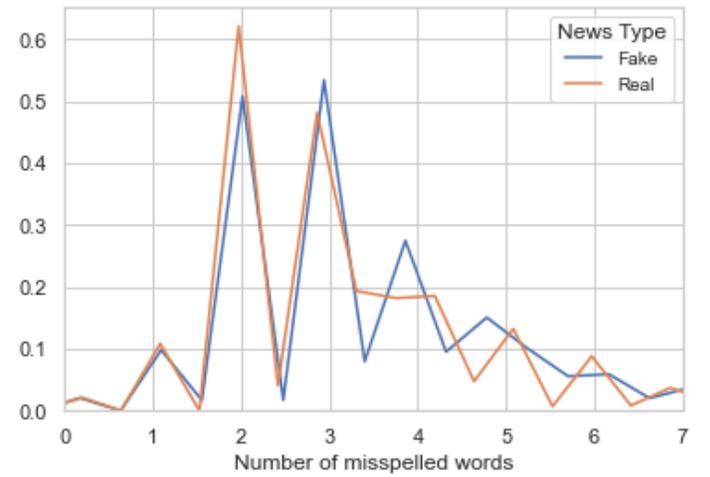

Fig.10

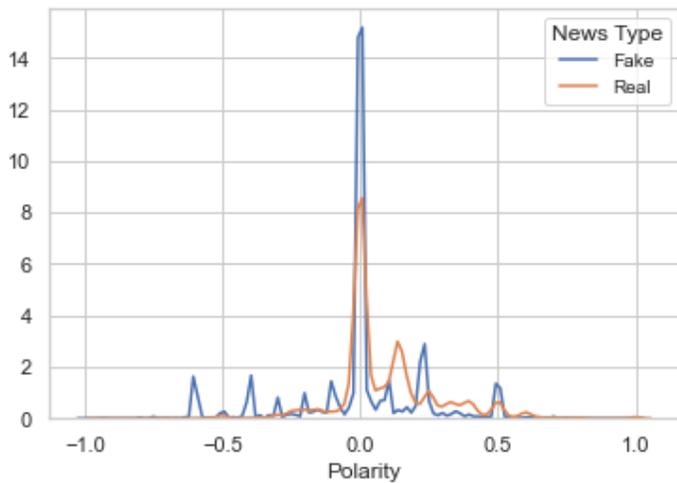

Fig. 11

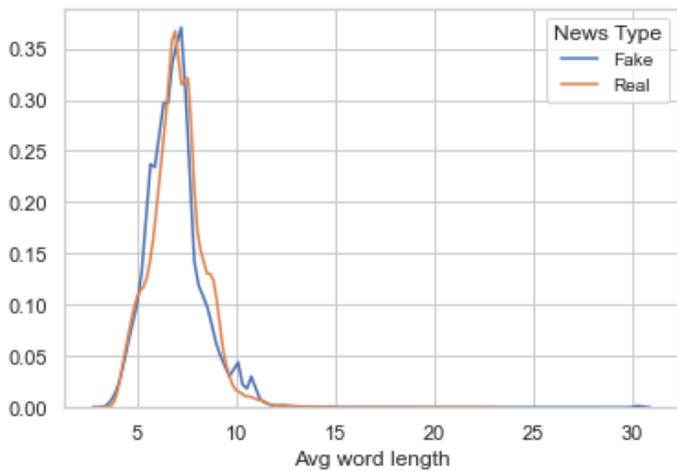

Fig. 12

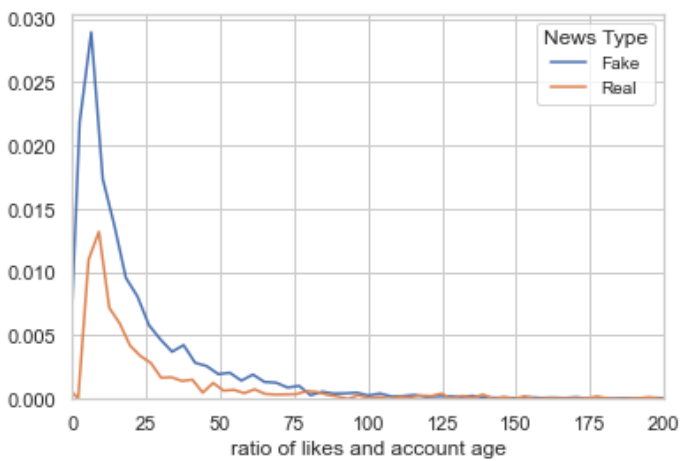

Fig. 13

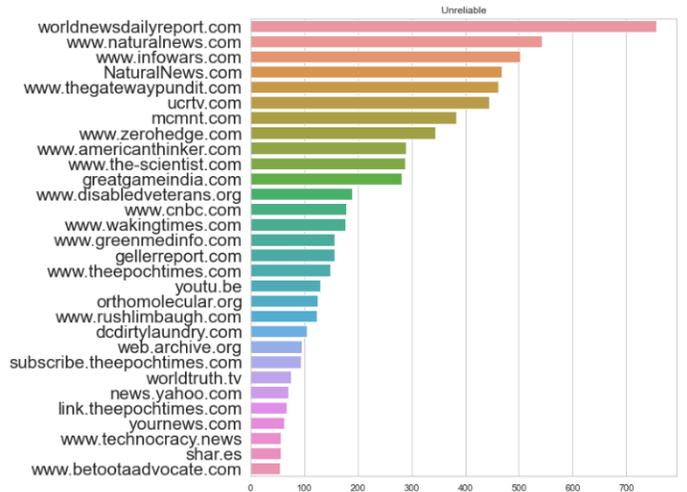

Fig.14

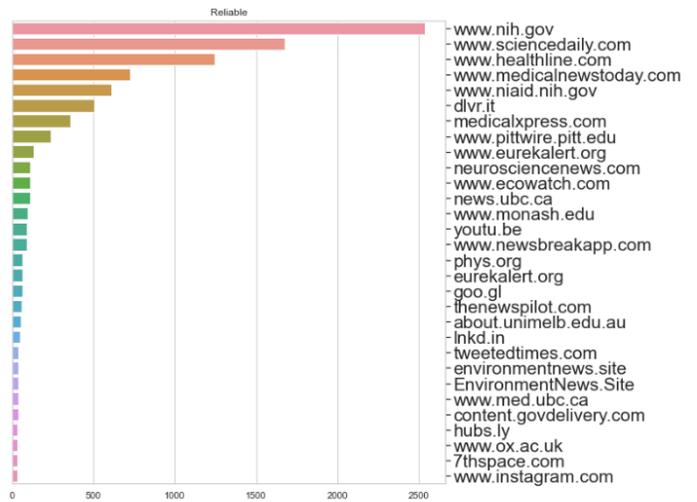

Fig. 15

In the metadata mined, we observed the presence of URLs in a number of tweet entities. People often use an external source to emphasize their point- be it malicious or just fact backing. A lot of these URLs are just clickbaits while others refer to sites that are well-known for spreading fake news. On the other hand, URLs to government or academic websites are generally thought to be trustworthy.

Fig.14 represents 20 most common websites in tweets spreading fake news. Therefore, we conclude that tweets containing content that refer to these sources can be deemed fake.On the other hand, Fig.15 refers to the twenty most common websites referred to in tweets that contain factual data.

Taking the above parameter into consideration,we introduced a new feature in the dataframe to tag if a particular tweet links to a bad or untrustworthy source. To study the impact of hashtags, we utilised word clouds. Fig.16 is the word cloud of hashtags related to tweets with misinformation while Fig.17 has hashtags associated with true information. One could assume that hashtags associated with misinformation can generate another parameter to identify if a tweet contains fake news or not. However, due to the large overlap between the two word clouds, this would not make a good feature.

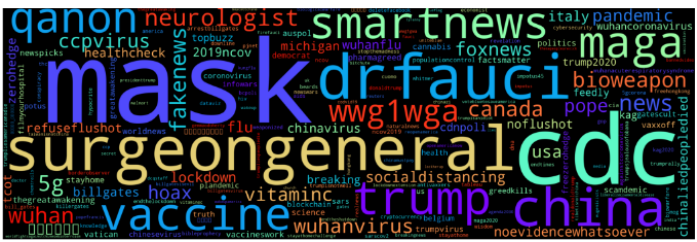
Fig.16

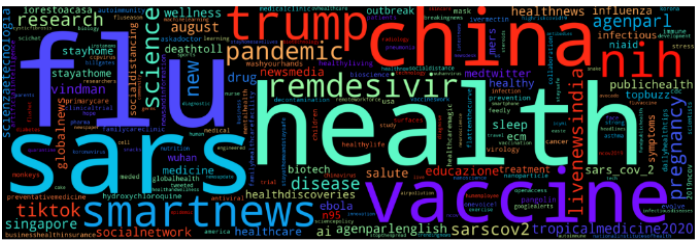
Fig.17

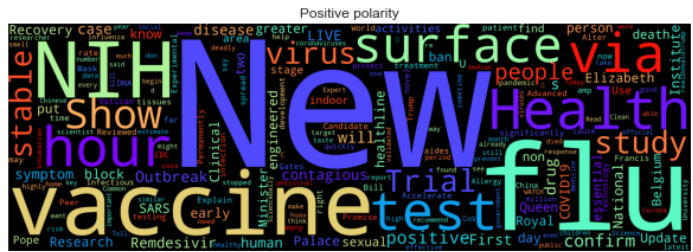

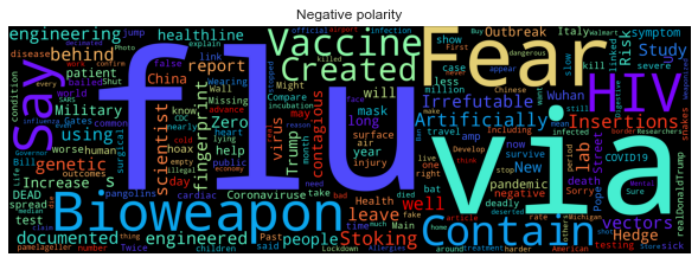

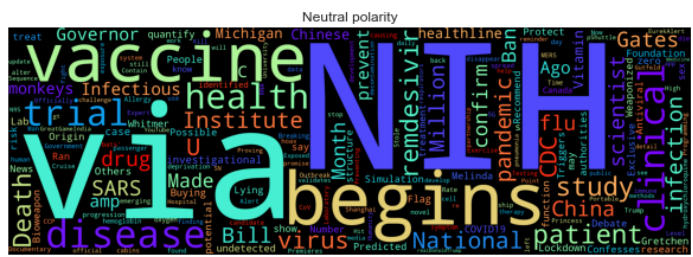
Fig. 18

Another metric to which could help measure feature importance is Jaccard similarity. It is defined as the size of the intersection divided by the size of the union of two label sets, is used to compare set of predicted labels for a sample to the corresponding set of labels in y_true.[18]

$$J(A, B) = \frac{|A \cap B|}{|A \cup B|} = \frac{|A \cap B|}{|A| + |B| - |A \cap B|}$$

If A and B are both empty, define J(A,B) = 1.
The Jaccard score of various features has been defined in Table 1

| Jaccard score | Feature | Conclusion |
| --- | --- | --- |
| 0.03258288 | Tweet contains sensitive information | Poor metric. Not a good feature. |
| 0.38654298 | User bio has a URL | Poor metric. Not a good feature. |
| 0.04850746 | User has default profile image | Poor metric. Not a good feature. |
| 0.01899539 | Tweet has hashtag(s) associated with misinformation | Poor metric. Not a good feature. |
| 0.30370768 | Tweet contains misspelled words | Poor metric. Not a good feature. |
| 0.709690113 | Tweet refers to untrustworthy source | Good metric. Illustrated above |

Table 1

Considering one can post on Twitter via numerous ways, the source of the tweet or the platform the user is posting from can be taken into account. Tweet source labels help to better understand how a Tweet was posted. This additional information provides context about the Tweet and its user.
However, since there is not a large difference between these label sources, it does not yield much useful information.

For misinformation, we observed that the ten most popular posting platforms were:

| Platform | Count of fake tweets |
| --- | --- |
| Twitter for iPhone | 2679 |
| Twitter Web Client | 2572 |
| Twitter Web App | 2245 |
| Twitter for Android | 1402 |
| Twitter for iPad | 504 |
| TweetDeck | 90 |
| WordPress | 88 |
| dlvrit | 53 |
| Hootsuite Inc | 52 |
| Twittimer | 41 |

Table 2

For tweets with real news, the ten most popular observed sources were::

| Platform | Count of real news tweets |
|---|---|
| Twitter Web Client | 3787 |
| Twitter for iPhone | 2705 |
| Twitter Web App | 2230 |
| Twitter for Android | 1543 |
| dlvrit | 547 |
| Twitter for iPad | 517 |
| Hootsuite Inc | 407 |
| WordPress | 343 |
| Buffer | 283 |
| TweetDeck | 278 |

Table 3

## IV. RESULT

Vectorization is the process of converting a group of text documents into numerical data. The process consists of two stages: first tokenize the string and assign an integer identifier for each possible token, followed by weighting of the token or condition to represent the importance of each token .In this project, Term Frequency Inverted Document Frequency (TF-IDF).was used for the same purpose. TF-IDF is used in machine learning and text mining as a weighting factor for features.The weight increases as the frequency of the word in the document increases but that is offset by how many times the word appears in the data set .This mechanism helps to eliminate the importance of very common words that are frequent across all documents and to consider words that rarely appear in the entire dataset. High TF-IDF weight is reached when a word has high TF in any given tweet and low DF of the word in the entire dataset.[19]

We considered supervised learning algorithms like the Random Forest classifier, AdaBoost Classifier, Bagging Classifier, Linear Support Vector Classification to model our problem and received the best results with the Random Forest algorithm. Random Forest[20] is a classifier containing several decision trees for different subsets of a given dataset and takes an average to improve the predictive accuracy of this dataset. Instead of relying on one decision tree, a random forest receives a prediction from each tree and based on the majority of votes predictions,predicts the final result.

Building upon the above definitions, we developed a TF-IDF classifier on the cleaned tweet text as the first step. Using TfidfVectorizer, top 2000 features with the highest TF score were chosen. The number 2000 was chosen arbitrarily to reduce the risk of lengthy vectors. This allowed for the introduction of a feature vector in context to TF-IDF in the dataset. Taking this new feature into account along with the pre-processed features, the Random Forest Classifier proved to be the best classifier and gave an f-1 score of 0.97621. The results obtained with other classifiers have been documented in Table4.

| Classifier | Precision | Recall | f1-score | Accuracy |
|---|---|---|---|---|
| **Random forest** | 0.980 | 0.978 | **0.976** | **0.978** |
| LinearSVC | 0.962 | 0.965 | 0.956 | 0.959 |
| Decision tree | 0.973 | 0.970 | 0.967 | 0.968 |
| AdaBoost | 0.963 | 0.984 | 0.973 | 0.972 |
| Bagging Classifer | 0.966 | 0.976 | 0.972 | 0.974 |

Table 4

## V. CONCLUSION

We have thus proved that we can detect misinformation on Twitter by applying Natural Language Processing techniques with simple Machine Learning algorithms. Apart from this, we analyzed the dataset, obtained the base attributes and mined more features to find out which parameters play an important role in the spread of misinformation and how different characteristics can be used to distinguish between truth and fiction. Social parameters like retweets/likes, semantic information, embedded URLs, author account activity and tweet content can be used to categorize if a tweet contains misinformation or not.